\documentclass{article}
\usepackage{spconf,amsmath,graphicx}

\usepackage{mwe} 
\usepackage[colorlinks,linkcolor=blue]{hyperref}
\usepackage{makecell}
\usepackage{multirow}
\usepackage{mathrsfs}
\usepackage{stfloats}

\title{Generalize Ultrasound Image Segmentation via Instant and Plug \& Play Style Transfer}
\name{
	\parbox{\linewidth}{\centering Zhendong Liu$^{1,2}$\sthanks{Authors contributed equally.}, Xiaoqiong Huang$^{1,2\ast}$, Xin Yang$^{1,2}$, Rui Gao$^{1,2}$, Rui Li$^{1,2}$, Yuanji Zhang$^{3}$, Yankai Huang$^{3}$, Guangquan Zhou$^{4}$, Yi Xiong$^{3}$, Alejandro F Frangi$^{1,5,6}$, Dong Ni$^{1,2}$ \sthanks{Corresponding author: nidong@szu.edu.cn.}}
}

\address{$^{1}$School of Biomedical Engineering, Health Science Center, Shenzhen University, China
\\$^{2}$ Medical UltraSound Image Computing (MUSIC) Lab, Shenzhen University, China
\\$^{3}$Department of Ultrasound, Luohu People’s Hospital, Shenzhen, China
\\$^{4}$ School of Biological Sciences and Medical Engineering, Southeast University, China
\\$^{5}$   Centre for Computational Imaging and Simulation Technologies in Biomedicine (CISTIB), \\
 School of Computing, University of Leeds, Leeds, UK
\\$^{6}$  Medical Imaging Research Center (MIRC), University Hospital Gasthuisberg,\\
Electrical Engineering Department, KU Leuven, Leuven, Belgium} 

\begin{document}
\maketitle
\begin{abstract}
Deep segmentation models that generalize to images with unknown appearance are important for real\-world medical image analysis. Retraining models leads to high latency and complex pipelines, which are impractical in clinical settings. The situation becomes more severe for ultrasound image analysis because of their large appearance shifts. In this paper, we propose a novel method for robust segmentation under unknown appearance shifts. Our contribution is three\-fold. First, we advance a one\-stage plug\-and\-play solution by embedding hierarchical style transfer units into a segmentation architecture. Our solution can remove appearance shifts and perform segmentation simultaneously. Second, we adopt Dynamic Instance Normalization to conduct precise and dynamic style transfer in a learnable manner, rather than previously fixed style normalization. Third, our solution is fast and lightweight for routine clinical adoption. Given 400 $\times$ 400 image input, our solution only needs an additional 0.2 ms and 1.92M FLOPs to handle appearance shifts compared to the baseline pipeline. Extensive experiments are conducted on a large dataset from three vendors demonstrate our proposed method enhances the robustness of deep segmentation models.
\end{abstract}

\section{Introduction}
\label{sec:intro}

\begin{figure}
	\centering
	\includegraphics[width=0.48\textwidth]{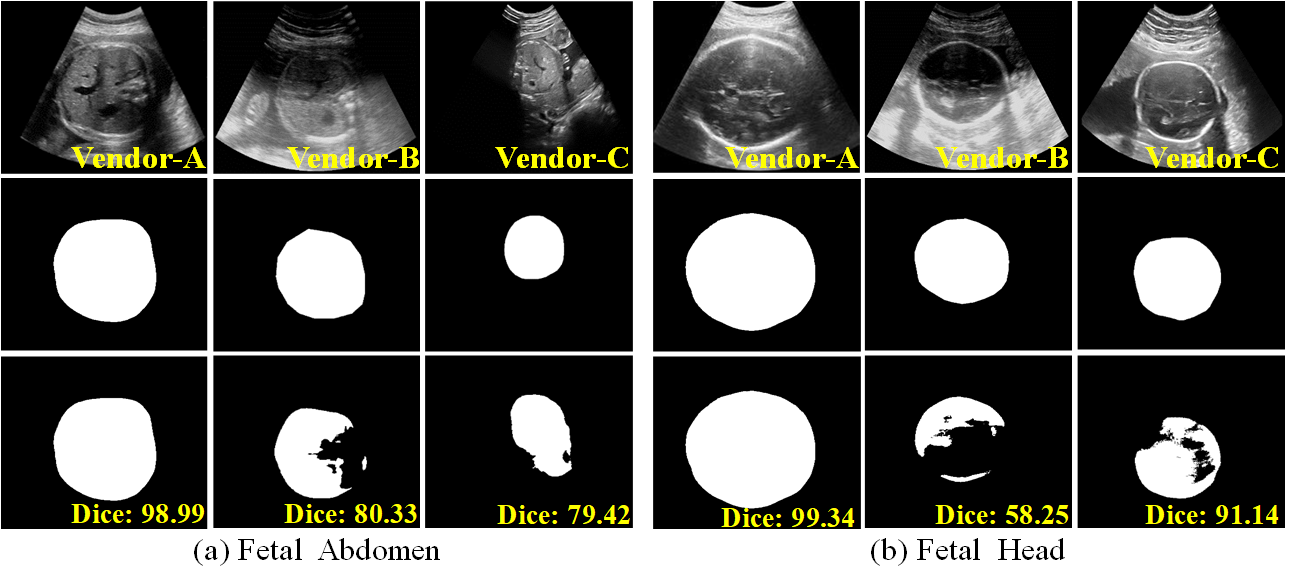}
	\caption{Illustration of segmentation degradation on (a) fetal abdomen and (b) fetal head ultrasound images acquired from different vendors. The second row shows the segmentation ground truth, and the third row shows the predictions by a deep model trained on Vendor-A.}
	\label{shift}
\end{figure}

The tremendous success of deep neural networks (DNNs) has benefitted medical image analysis \cite{liu2019deep}. However, deployment of DNN models in real clinical scenarios is threatened by appearance shifts that degrade their performance. Different sources of appearance variation affect routine medical image acquisition, including operators, protocols, vendors, parameters and tissue properties, all of which can lead to unpredictable image appearance changes  \cite{gibson2018inter,yan2019edge}. The adverse effect of appearance shift on ultrasound image segmentation can be observed in Fig.\ref{shift}. Making DNNs robust against appearance shift is along the last mile before they can be clinically adopted. \par

Automated solutions to remove image appearance shift are highly desirable but challenging. Real clinical applications often require the solution to be fast, practical and straightforward. Whereas, limited computation resources, the impossibility of retraining, and unpredictable appearance shifts are stringent constraints faced by these solutions, especially for ultrasound image analysis. Recently, Domain Adaptation (DA) has been proposed for dealing with image appearance shifts. Aligning appearance spaces \cite{nie2018medical,yang2018generalizing} or feature spaces \cite{kamnitsas2017unsupervised,huo2017adversarial,zhang2018translating} of different domains were explored using generative adversarial learning. Although DA is attractive, it depends heavily on sufficient data from the target domain for complex retraining. The need to find a source-target domain mapping confines these solutions to cases where both domains are defined a priori. By revisiting the basic definition of appearance shift, style transfer \cite{gatys2016image} (ST) inspires a new and intuitive way to tackle this problem. ST removes appearance shift by rendering the appearance of the target content image as a reference style image. Compared to DA, ST is independent on the target domain, retraining-free, and suitable for images with unknown appearance shift. Ma \textit{et al.} \cite{ma2019neural} made the early attempt to exploit an online ST to reduce the appearance variation for better cardiac MR segmentation. Liu \textit{et al.} \cite{liu2020remove} proposed an Adaptive Instance Normalization (AdaIN) based style transfer module for vendor adaptation. However, these methods only regard ST as a pre-processing module isolated from segmentation model. This two-stage setting not only consumes extra computation resources but also blocks ST-segmentation interaction.\par

In this paper, we propose a novel ST based one-stage framework for robust ultrasound image segmentation against unknown appearance shift. Our contribution is three-fold. \textit{First}, with the plug-and-play design of the ST module, we unify the ST and segmentation model into a one-stage framework to remove unknown appearance shift and perform segmentation simultaneously. \textit{Second}, we adopt the Dynamic Instance Normalization (DIN) \cite{jing2019dynamic} to conduct style transfer at multiple layers of segmentation model through a learnable manner, which provides precise affine parameters for more accurate style transfer. \textit{Third}, our solution is lightweight and reduces the computation burden for better deployment in clinical scenarios. Given a 400$\times$400 image input, our solution only needs 0.2\textit{ms} and 1.92M FLOPs to handle appearance shift. Segmentation models following our solution need no samples for time-consuming retraining before they can take domain-unknown images. Extensive experiments are conducted on 6,532 fetal head (FH) and abdomen (FA) ultrasound images from three vendors. The results show our proposed method outperforms state-of-the-art methods. \par
	
\begin{figure}
	\centering
	\includegraphics[width=0.48\textwidth]{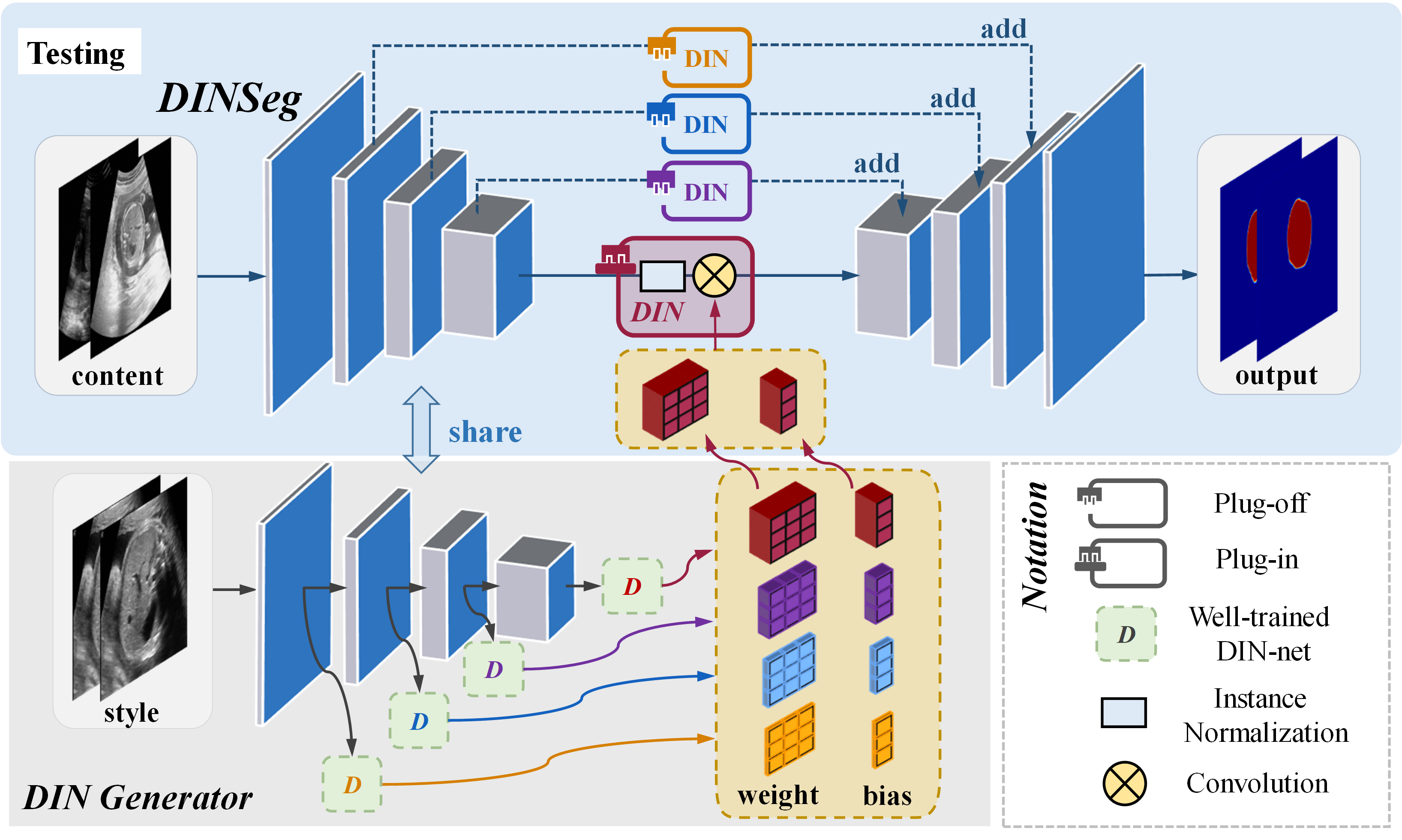}
	\caption{Schematic view of our proposed framework.}
	\label{overall_model}
\end{figure}

\section{Methodology}
\label{sec:Methodology}

Figure \ref{overall_model} presents a schematic view of our method. We first train the segmentation model on Vendor A (style) images by using a U-net architecture \cite{ronneberger2015u}, which achieved remarkable success especially in medical image segmentation. We then freeze this model to train hierarchical DIN-nets by only using its encoder part and construct the final plug-and-play model (DINSeg) for segmenting content images from other vendors. \par

\subsection{One-stage Plug \& Play Framework}
Real clinical scenarios hold tough challenges for DNNs to combat appearance shift. Recently, ST has been proposed to remove the gap by rendering the appearance of unknown content images as style images. Previous ST-based methods \cite{ma2019neural,liu2020remove} trained two models and developed two-stage systems for segmenting images with appearance shift. In the first stage, the content image is transferred into a stylized image by the ST model. In the second stage, the segmentation model trained on style images is performed on the stylized image and get the result. This two-stage setting is straightforward, but it does not only consume extra computation resources but also blocks the interplay between ST and segmentation. \par

As shown in Fig. \ref{overall_model}, we propose to simplify the pipeline into a novel one-stage design. Instead of a separate model, the DIN unit serves as the plug-and-play module for segmentation. Specifically, layer-specific DIN units are extracted from the trained DIN generator and then can be plugged into the segmentation model. These units parameterized by the appearance of a style image can align statistics of the content feature with those of the style features. In this plug-and-play setting, the choice of ST units is critical. \par

\subsection{Choice of Style Transfer Units}
Huang \textit{et al.} \cite{huang2017arbitrary} observed that matching feature statistics can achieve arbitrary style conversion. They proposed AdaIN to align style statistics after performing\textit{ Instance Normalization} (IN) on feature maps of the content image. AdaIN is powerful, but the affine parameters for style transfer are empirically defined as the basic statistics of mean and standard deviation, which may lead to the suboptimal ST performance. We propose to introduce \textit{Dynamic Instance Normalization} (DIN) \cite{jing2019dynamic} to generate required plug-and-play DIN units for the first time in the literature. DIN units are developed in a learnable way and thus can encode a sophisticated style patterns. \par

\begin{figure}
	\centering
	\includegraphics[width=0.48\textwidth]{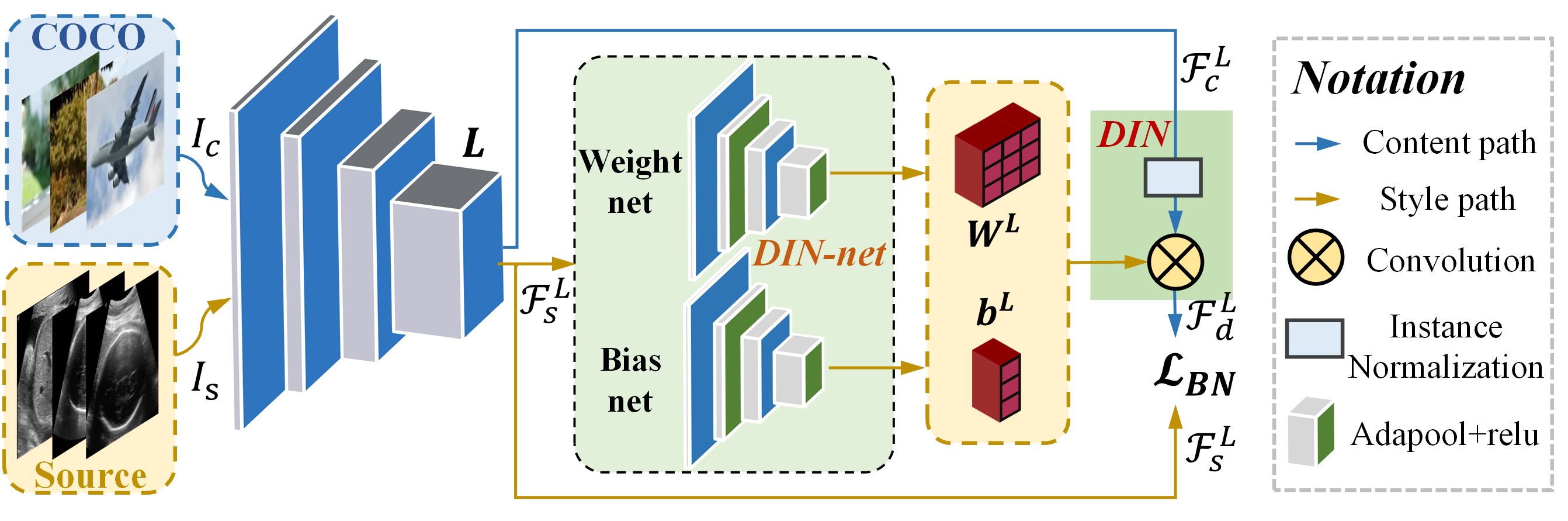}
	\caption{System design to train DIN-net. The weight $W^L$ and bias $b^L$ are the affine parameters of DIN unit at layer $L$.}
	\label{din}
\end{figure}

As shown in Fig.~\ref{din}, given a style image $I_s$, the affine parameters $W^L$ and $b^L$ are learned from the feature maps $\mathcal{F}_{s}^{L}$ of $I_s$ by \textit{DIN-net}. Then, normalized $\mathcal{F}_{c}^{L}$ is stylized by dynamic convolution with the learned weight $W^L$ and bias $b^L$. DIN operation can be formulated as:
\begin{equation}
{\rm \textbf{DIN}} (\mathcal{F}_{c}^{L})= {\rm \textbf{IN}}(\mathcal{F}_{c}^{L})\otimes W^L + b^L 
\label{din_eq}
\end{equation}

\subsection{Training of Dynamic Instance Normalization Unit}
Figure~\ref{din} depicts the details of training DIN-net. Unlike \cite{jing2019dynamic}, which uses an encoder-decoder architecture to train DIN-net for ST, we only keep the encoder as our system core to simplify the training, which will generate the plug-and-play units. This modification also accelerates and stabilizes the training process. Specifically, our system shares the same encoder as the segmentation model. Only the DIN-net, including weight and bias branches, need to be trained. These two branches consist of two convolutional and adaptive pooling operations for handling arbitrary input sizes. To provide enough instances to improve and verify the ST performance, we use a large natural images corpus (COCO dataset \cite{lin2014microsoft}) as content image inputs for DIN-net training. With the training, DIN-net can learn complex and rich style patterns.

As shown in Fig. \ref{din}, ultrasound images serve as style images $I_s$. The frozen encoder extracts the style feature maps $\mathcal{F}_{s}^{L}$ at layer $L$. A layer-specific DIN-net is attached at layer $L$. The weight net and bias net are trained and then generate the affine parameters of DIN unit at layer $L$. Suggested by Li \textit{et al.} \cite{li2017demystifying}, measuring the alignment loss via Batch Normalization (BN) statistics can achieve style transfer. Hence, given $\mu$ and $\sigma$ as the channel-wise mean and standard deviation, we minimize the BN-Statistics Matching loss to convey the style information to DIN output feature $\mathcal{F}_{d}^{L}$ gradually as follow:
\begin{equation}
\mathcal{L}_{BN} = {\Vert\mu(\mathcal{F}_{d}^{L})-\mu(\mathcal{F}_{s}^{L})\Vert}_2+{\Vert\sigma(\mathcal{F}_{d}^{L})-\sigma(\mathcal{F}_{s}^{L})\Vert}_2
\end{equation}

\section{Experimental Results}
\label{sec:Experimental Results}
\subsection{Dataset and Training Details.}
Our datasets consist of the fetal head (FH), and abdomen (FA) images acquired from three ultrasound vendors, denoted as Ven-A, Ven-B and Ven-C. Images from Ven-B and Ven-C were further adjusted with different time gain compensations (TGC) to mimic more complex appearance shift variations. Our experiments involved 4200, 1516 and 816 ultrasound images from Ven-A, Ven-B and Ven-C, respectively. Experienced experts provided ground truth for all images. Ven-A images were set as style images to train the segmentation network. In contrast, images from the other two vendors were set as the unseen content images with unknown appearance shift. All the data acquisition was approved by local IRB and anonymized for segmentation. \par

We trained DIN-nets using 10,000 images from Microsoft COCO dataset as content images and 200 images from Ven-A as style images. We noted that DIN-net learns the affine parameters from ultrasound image in Vendor-A (style), rather than COCO content images. The COCO dataset we adopted can simulate complex style distribution and enable DIN-net to learn richer style patterns. Moreover, it is challenging to collect massive amount of ultrasound images to train DIN-nets fully but versatile COCO datasets are easy to obtain. Adam optimizer with learning rate $10^{-3}$ is set to minimize $\mathcal{L}_{BN}$. We trained a typical U-net segmentation model on Ven-A images and froze it during testing (baseline). We conducted all experiments with PyTorch on an NVIDIA GTX 2080 Ti GPU. \par

\subsection{Quantitative and Qualitative Evaluation.}
Based on our framework and the usage of DIN plug-ins, we devised two variants of our methods, namely Single-DINSeg (\textit{S-DINSeg}) and Multi-DINSeg (\textit{M-DINSeg}). As depicted in Fig.~\ref{overall_model}, \textit{S-DINSeg} only plugs a single DIN at the end of the segmentation encoder, while \textit{M-DINSeg} has an extra DIN in each of the last three skip connections. We compared them with the solution replacing DIN with AdaIN \cite{huang2017arbitrary}, and got two variants, \textit{S-AdaINSeg} and \textit{M-AdaINSeg}. We also implemented two typical two-stage frameworks for comparison, including the \textit{StyleSegor} \cite{ma2019neural} and \textit{WaveCT-AIN} (WCT-AIN) \cite{liu2020remove}. GAN based DA methods are not considered for comparison in this work, since they need samples from Ven-B and Ven-C for retraining. We used in total 6 indicators for evaluation, including Dice coefficient (Dice), Jaccard index (JAC), Hausdorff Distance of Boundaries (HDB), Average Surface Distance (ASD), Precision (PRE) and Recall (REC).

\begin{table}[htbp]
	\caption{Quantitative evaluation across different vendors.}
	\scalebox{0.82}{
		\begin{tabular}{p{0.9cm}p{1.8cm}|p{1.8cm}p{1.8cm}p{1.8cm}}
			\hline{}
			& \centering\multirow{2}*{\textbf{Metric}} &\makecell[c]{\textbf{Baseline}} & \makecell[c]{\textbf{S-DINSeg}} & \makecell[c]{\textbf{M-DINSeg}} \\
			\cline{3-5}
			~ & & \makecell[c]{FA/FH} &   \makecell[c]{FA/FH} &   \makecell[c]{FA/FH} \\
			\hline
			
			\centering\multirow{2}*{Ven-B}&\centering\textbf{Dice(\%)}  & \makecell[c]{88.63/94.93} & \makecell[c]{92.27/96.20} & \makecell[c]{93.58/96.97} \\
			
			&\centering\textbf{HDB(pixel)} & \makecell[c]{17.03/11.87} & \makecell[c]{13.39/7.52} & \makecell[c]{12.50/6.07}\\
			\hline
			\centering\multirow{2}*{\makecell[c]{{Ven-C}}} &\centering\textbf{Dice(\%)} & \makecell[c]{94.69/95.69} & \makecell[c]{95.13/96.27} & \makecell[c]{95.00/96.97} \\
			&\centering\textbf{HDB(pixel)} & \makecell[c]{17.77/11.21} & \makecell[c]{10.05/8.45} & \makecell[c]{10.34/7.31} \\
			\hline
	\end{tabular}}
	\label{compare_self}
	
\end{table}

\begin{figure*}

	\centering
	\includegraphics[scale=0.5]{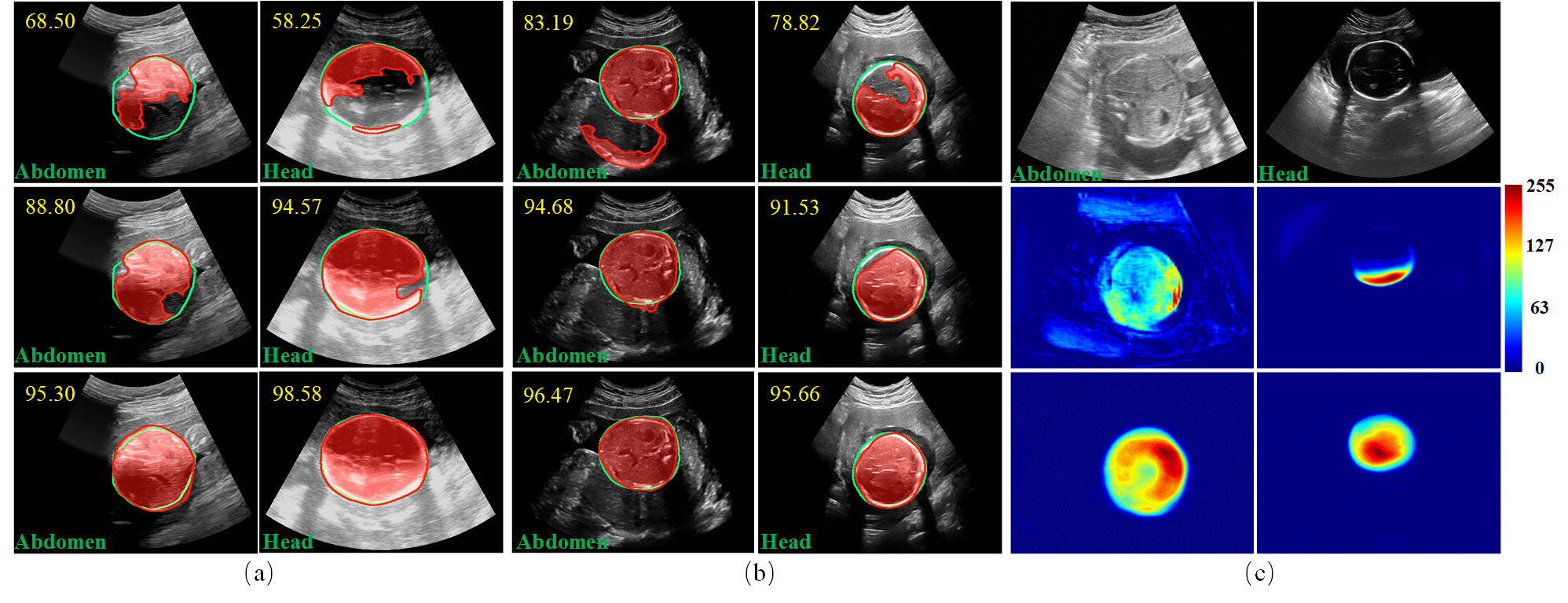}
	\caption{(a)-(b) Segmentation results on Ven-B and Ven-C FA (left) and FH (right) images. Baseline, S-DINSeg and M-DINSeg results are listed from top to bottom. Green curve, red area and yellow digits denote ground truth, segmentation and Dice, respectively. (c) Feature maps before (middle row) and after (bottom row) DIN.}
	\label{seg_res}

\end{figure*}

Table \ref{compare_self} shows the segmentation results of fetal images from two vendors. Both S-DINSeg and M-DINSeg get consistent improvements on two datasets, even when the baseline on Ven-C is already competitive. This indicates the efficacy of our proposed framework against unknown appearance shift. With the DIN used in multiple sites of the segmentation network, the results of M-DINSeg demonstrates that hierarchical DIN units can remove appearance shift better by encoding rich style information of different levels.

\begin{table}[htbp]

	\centering
	\caption{Quantitative comparisons among different methods.}
	\scalebox{0.57}{
		\begin{tabular}{c|cccccc}
			\hline
			\multirow{2}*{\textbf{Methods}} & \textbf{Dice(\%)} & \textbf{JAC(\%)} & \textbf{HDB} & \textbf{ASD} & \textbf{PRE(\%)} & \textbf{REC(\%)} \\
			\cline{2-7}
			&\makecell[c]{FA/FH} & \makecell[c]{FA/FH} & \makecell[c]{FA/FH} &  \makecell[c]{FA/FH}& \makecell[c]{FA/FH}&\makecell[c]{FA/FH} \\
			\hline
			Baseline & 88.63/94.93 & 83.42/91.67 & 17.03/11.87 & 4.11/2.86 & 95.35/97.29 & 86.11/94.03 \\
			HistEqual & 84.16/87.69 & 78.68/83.54 & 20.47/14.94 & 4.63/3.15 & 93.93/93.59 & 81.21/85.86\\
			StyleSegor \cite{ma2019neural} & 89.89/95.11 & 84.54/91.74 & 16.88/11.78 & 3.97/2.81 & 95.33/96.56 & 88.12/94.42 \\
			WCT-AIN \cite{liu2020remove} & 91.87/96.21 & 86.82/93.10 & 14.76/7.34 & 3.71/2.11 & 96.36/97.13 & 89.63/95.33 \\
			S-AdaINSeg & 91.65/96.24 & 86.63/93.41 & 14.17/7.30 & 3.74/2.07 & 96.20/97.30 & 89.70/95.84 \\
			M-AdaINSeg & 92.52/96.42 & 87.71/93.64 & 13.33/7.22 & 3.52/2.02 & 95.77/97.00 & 91.26/96.37 \\
			\hline
			\textbf{S-DINSeg} & 92.27/96.20 & 86.85/93.03 & 13.39/7.52 & 3.85/2.17 & \textbf{96.58/97.54} & 89.77/95.19 \\
			\textbf{M-DINSeg} & \textbf{93.58/96.97} & \textbf{88.76/94.38} & \textbf{12.50/6.07} & \textbf{3.52/1.76} & 95.80/96.70 & \textbf{92.42/97.43} \\
			\hline
	\end{tabular}}
	\label{compare_method}
\end{table}

Table \ref{compare_method} lists the quantitative comparisons among our methods and other ST-based methods on Ven-B images, due to the space limitation. This indicates the efficacy of ST for segmenting images with appearance shift. Best results are achieved by our proposed method. M-DINSeg improves the Dice index by about 5 percent for FA and 2 percent for FH over their baselines. Among these methods, one-stage solutions, including M-AdaINSeg and DINSeg, show consistent advantages over the two-stage frameworks, like the StyleSegor and WCT-AIN. Hence, simplifying the pipeline and directly conducting ST in segmentation network are tractable and superior in removing appearance shift.

Figure \ref{seg_res} (a-b) visualize the results of our methods on segmenting FA and FH from Ven-B and Ven-C, respectively. With DIN plug-in, S-DINSeg can almost recover the segmentation from the broken masks obtained by the baseline. When DIN units further plug into the skip connections, M-DINSeg presents the most visually plausible segmentation and highest Dice when compared to the ground truth. Fig. \ref{seg_res} (c) gives insights into the change in feature maps before and after DIN-based style transfer. DIN removes appearance shift and makes the segmentation get strong activations only around regions of interest.

\begin{table}
	\centering
	\caption{Model complexity evaluation.}
	\scalebox{0.72}{
		\begin{tabular}{ccc|cc|cc}
			\hline
			\multirow{2}*{\textbf{Methods}} & \multicolumn{2}{c}{\textbf{FLOPs(G)}} & \multicolumn{2}{c}{\textbf{Params(M)}} & \multicolumn{2}{c}{\textbf{Time(ms)}} \\
			\cline{2-7}
			& Transfer & Whole & Transfer & Whole & Transfer & Whole \\
			\hline
			StyleSegor \cite{ma2019neural} & 49.17 & 209.12 & 138.36 & 172.89 & 3000 & 3029.88 \\
			WCT-AIN \cite{liu2020remove}     & 70.77 & 230.72 & 10.66  & 45.19  & 75   & 104.98 \\
			S-AdaINSeg & 6.4e{-4} & 159.95 & / & 34.53 & 0.17 & 30.15 \\
			M-AdaINSeg & 9.81e{-3} & 159.95 & / & 34.53 & 0.88 & 30.86 \\
			\hline
			\textbf{S-DINSeg} & 1.92e{-3} & 159.95 & 0.002 & 34.532 & 0.20 & 30.26 \\
			\textbf{M-DINSeg} & 1.94e{-2} & 159.95 & 0.004 & 34.534 & 1.99 & 31.87 \\
			\hline
	\end{tabular}}
	\label{model_complexity}
\end{table}

Table \ref{model_complexity} investigates the computation complexity of different methods (input size is 400$\times$400). The number of floating-point operations (FLOPs), total parameters (Params) and inference time (ms) are reported criteria. Using the same segmentation network, we evaluated the time of ST (Transfer) and the whole pipeline (Whole) separately. Results show that our system based on one-stage plug-ins solution enhance segmentation robustness in terms of efficiency and complexity.

\section{Conclusion}
\label{sec:Conclusion}

In this work, we propose \textit{DINSeg}, which unifies DIN and segmentation network in a single model to increase both generalization capacity and inference efficiency. We show that \textit{DINSeg} achieves consistent improvement over existing effective systems on generalizing deep model across a new domain. The proposed system is high-speed and lightweight being thus amenable for deployment in clinical settings.

\section{Acknowledgments}
This work was supported by the grant from National Key R\&D Program of China (No.2019YFC0118300), Shenzhen Peacock Plan (No. KQTD2016053112051497, KQJSCX201\\80328095606003).

\section{Compliance with Ethical Standards}
\label{Compliance with Ethical Standards}
We state that our work is an academic study for which no ethical approval was required.


	
\bibliographystyle{IEEEbib}
\bibliography{refs}

\begin{thebibliography}{10}

\bibitem{liu2019deep}
S.~Liu, Y.~Wang, X.~Yang, B.~Lei, L.~Liu, S.~Li, D.~Ni, and T.~Wang,
\newblock ``Deep learning in medical ultrasound analysis: a review,''
\newblock {\em Engineering}, 2019.

\bibitem{gibson2018inter}
E.~Gibson, Y.~Hu, N.~Ghavami, H.~U Ahmed, C.~Moore, M.~Emberton, H.~J Huisman,
  and D.~C Barratt,
\newblock ``Inter-site variability in prostate segmentation accuracy using deep
  learning,''
\newblock in {\em International Conference on Medical Image Computing and
  Computer-Assisted Intervention}. Springer, 2018, pp. 506--514.

\bibitem{yan2019edge}
W.~Yan, Y.~Wang, M.~Xia, and Q~Tao,
\newblock ``Edge-guided output adaptor: Highly efficient adaptation module for
  cross-vendor medical image segmentation,''
\newblock {\em IEEE Signal Processing Letters}, vol. 26, no. 11, pp.
  1593--1597, 2019.

\bibitem{nie2018medical}
D.~Nie, R.~Trullo, J.~Lian, L.~Wang, C.~Petitjean, S.~Ruan, Q.~Wang, and
  D.~Shen,
\newblock ``Medical image synthesis with deep convolutional adversarial
  networks,''
\newblock {\em IEEE TBME}, 2018.

\bibitem{yang2018generalizing}
X.~Yang, H.~Dou, R.~Li, X.~Wang, C.~Bian, S.~Li, D~Ni, and P.A. Heng,
\newblock ``Generalizing deep models for ultrasound image segmentation,''
\newblock in {\em International Conference on Medical Image Computing and
  Computer-Assisted Intervention}. Springer, 2018, pp. 497--505.

\bibitem{kamnitsas2017unsupervised}
K.~Kamnitsas, C.~Baumgartner, C.~Ledig, V.~Newcombe, J.~Simpson, A.~Kane,
  D.~Menon, A.~Nori, A.~Criminisi, D.~Rueckert, et~al.,
\newblock ``Unsupervised domain adaptation in brain lesion segmentation with
  adversarial networks,''
\newblock in {\em International conference on information processing in medical
  imaging}. Springer, 2017, pp. 597--609.

\bibitem{huo2017adversarial}
Y.~Huo, Z.~Xu, S.~Bao, A.~Assad, R.~G Abramson, and B.~A Landman,
\newblock ``Adversarial synthesis learning enables segmentation without target
  modality ground truth,''
\newblock in {\em ISBI}. IEEE, 2018, pp. 1217--1220.

\bibitem{zhang2018translating}
Z.~Zhang, L.~Yang, and Y.~Zheng,
\newblock ``Translating and segmenting multimodal medical volumes with
  cycle-and shapeconsistency generative adversarial network,''
\newblock in {\em CVPR}, 2018, pp. 9242--9251.

\bibitem{gatys2016image}
L.~A Gatys, A.~S Ecker, and M.~Bethge,
\newblock ``Image style transfer using convolutional neural networks,''
\newblock in {\em CVPR}, 2016, pp. 2414--2423.

\bibitem{ma2019neural}
C.~Ma, Z.~Ji, and M.~Gao,
\newblock ``Neural style transfer improves 3d cardiovascular mr image
  segmentation on inconsistent data,''
\newblock in {\em International Conference on Medical Image Computing and
  Computer-Assisted Intervention}. Springer, 2019, pp. 128--136.

\bibitem{liu2020remove}
Z.~Liu, X.~Yang, R.~Gao, S.~Liu, H.~Dou, S.~He, Y.~Huang, Y.~Huang, H.~Luo,
  Y.~Zhang, et~al.,
\newblock ``Remove appearance shift for ultrasound image segmentation via fast
  and universal style transfer,''
\newblock {\em arXiv preprint arXiv:2002.05844}, 2020.

\bibitem{jing2019dynamic}
Y.~Jing, X.~Liu, Y.~Ding, X.~Wang, E.~Ding, M.~Song, and S.~Wen,
\newblock ``Dynamic instance normalization for arbitrary style transfer,''
\newblock {\em arXiv preprint arXiv:1911.06953}, 2019.

\bibitem{ronneberger2015u}
O.~Ronneberger, P.~Fischer, and T.~Brox,
\newblock ``U-net: Convolutional networks for biomedical image segmentation,''
\newblock in {\em International Conference on Medical image computing and
  computer-assisted intervention}. Springer, 2015, pp. 234--241.

\bibitem{huang2017arbitrary}
X.~Huang and S.~Belongie,
\newblock ``Arbitrary style transfer in real-time with adaptive instance
  normalization,''
\newblock in {\em ICCV}, 2017, pp. 1501--1510.

\bibitem{lin2014microsoft}
T.Y. Lin, M.~Maire, S.~Belongie, J.~Hays, P.~Perona, D.~Ramanan, P.~Doll{\'a}r,
  and C~L. Zitnick,
\newblock ``Microsoft coco: Common objects in context,''
\newblock in {\em European conference on computer vision}. Springer, 2014, pp.
  740--755.

\bibitem{li2017demystifying}
Y.~Li, N.~Wang, J.~Liu, and X.~Hou,
\newblock ``Demystifying neural style transfer,''
\newblock {\em arXiv preprint arXiv:1701.01036}, 2017.

\end{thebibliography}
	
\end{document}